\documentclass[
reprint,
superscriptaddress,
amsmath,amssymb,
aps,
prb,
]{revtex4-2}

\usepackage{graphicx,import}
\usepackage{dcolumn}
\usepackage{upgreek}
\usepackage{physics}
\usepackage{amsmath}
\usepackage{setspace}
\usepackage{bm}
\usepackage[breaklinks=true,colorlinks=true,linkcolor=blue,urlcolor=blue,citecolor=blue]{hyperref}
\allowdisplaybreaks[4]
\UseRawInputEncoding

\begin{document}

	\preprint{APS/123-QED}
	
	\title{Perfect splitting of a two-photon pulse}
	
	\author{Mads M. Lund}
	\email{mml@phys.au.dk}
	\affiliation{Center for Complex Quantum Systems, Department of Physics and Astronomy, Aarhus University, 8000 Aarhus C, Denmark}
	
	\author{Fan Yang}
	\email{fanyang@phys.au.dk}
	\affiliation{Center for Complex Quantum Systems, Department of Physics and Astronomy, Aarhus University, 8000 Aarhus C, Denmark}
	
	\author{Klaus M{\o}lmer}
	\email{klaus.moelmer@nbi.ku.dk}
	\affiliation{Niels Bohr Institute, University of Copenhagen, Blegdamsvej 17, 2100 Copenhagen, Denmark}

	\newcommand{\hatb}{\hat{b}}
	\newcommand{\hata}{\hat{a}}
	\newcommand{\hatsigma}{\hat{\sigma}}

	
	\begin{abstract}
		
		 We employ a cascaded system approach to numerically simulate the interaction of photon pulses with a two-level scatterer in a chiral waveguide QED setup. It is possible to expand any pure state of two photons as a superposition of orthogonal two-photon pulses. We show that the scattering of a two-photon pulse of suitable duration may lead to an entangled output state between predominately two equally populated two-photon states. In a complementary wave packet basis, this state is a product state of two orthogonal single photon wave packets.  
		 The time reversal of the above scattering process allows for a perfect combination of distinguishable single-photon wave packet into a single-mode pulse carrying two identical photons.
		
	\end{abstract}
	
	\maketitle

\section{Introduction}
\begin{figure}[t]
    \includegraphics[width = 0.89\columnwidth]{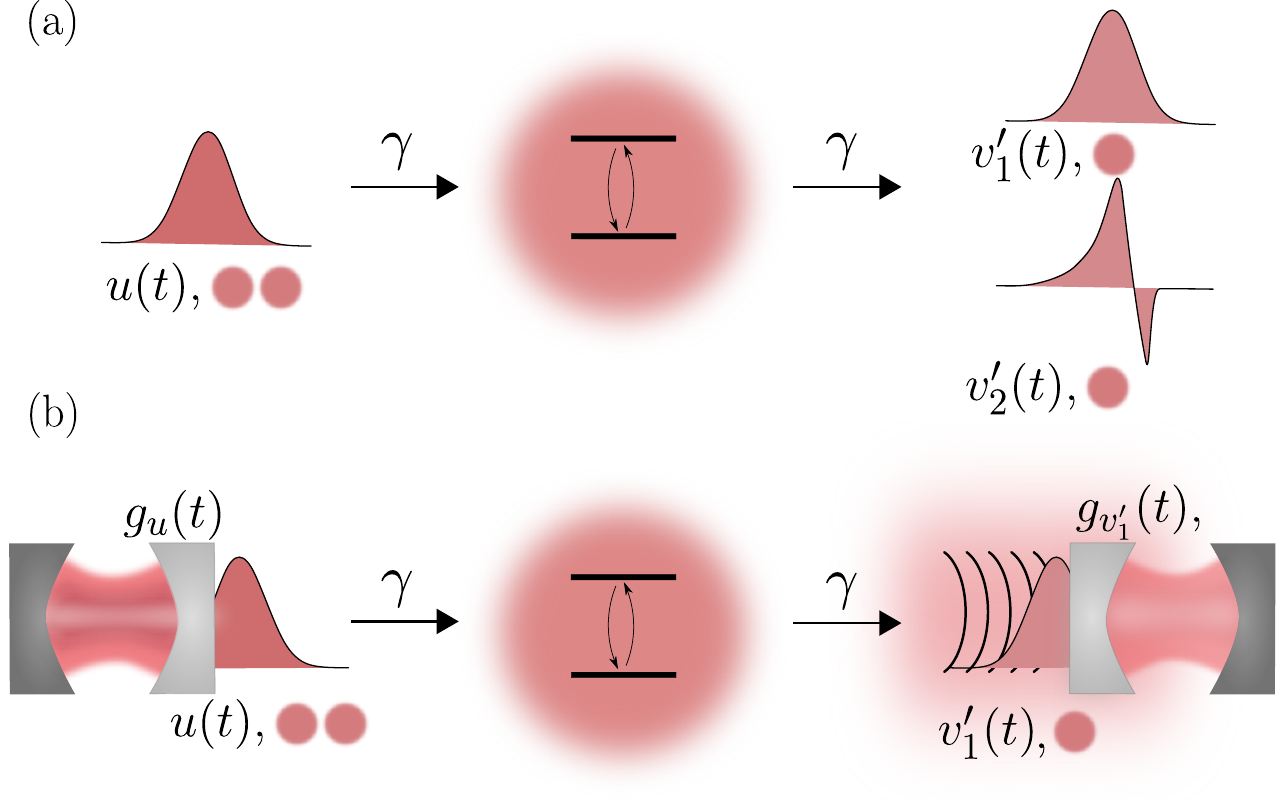}
    \caption{(a) Illustration of a single temporal mode, $u(t)$, occupied by a two-photon state. The photons scatters chirally on a two-level system, with a dipole interaction strength $\sqrt{\gamma}$, and forms an output field with two orthogonal modes, $v_1'(t)$ and $v_2'(t)$ [Eq.'s \eqref{eq:v_1'} and \eqref{eq:v_2'}], each occupying a single-photon Fock state. (b) Representation of the scattering of a single input wave packet and detection of a single output wave packet mode by virtual cavities. The leftmost virtual cavity, with time-dependent coupling, $g_u(t)$, in Eq. \eqref{eq:gut}, emits its initial two-photon Fock state content in the temporal mode $u(t)$. The rightmost cavity, with time-dependent coupling $g_{v_1'}(t)$ in Eq. \eqref{eq:gvt}, absorbs the quantum state content of the temporal mode, $v_1'(t)$, of the scattered field.}
    \label{fig:fig1}
    \end{figure}

Quantum states of light provide indispensable resources for quantum sensing \cite{gravwave1,Gravwave2,Schwartz2013,Wolfgramm2013} and quantum information processing \cite{Kimble2008,O'Brien2009}. In recent years large efforts have been devoted to their preparation, interactions with matter, and detection at the fundamental quantum level. Processes that can produce non-classical states of light and perform non-classical transformations of these states play a key role in this endeavor. Such processes put strict requirements on the strength, coherence, and bandwidth of the interactions applied. In cavity QED, the interaction between a single optical mode and a classically driven two-level system (TLS) displays strong non-linearity and permits the construction of a variety of field quantum states \cite{RevModPhys.73.565,PhysRevLett.57.2520}. For traveling fields the situation is generally more complex, as non-linear interactions may lead to the population of many propagating modes, reducing the purity of the desired few-mode states \cite{Raymer_2020,PhysRevA.99.023857,PhysRevLett.84.5304,PhysRevA.99.023857,Carmichael_cas,gardiner_cas,baragiola2012,baragiola2017,fischer2018scattering,PRX.5.041017,qintwpor,inoutqpulse}. 
An effective theoretical description was recently provided for the interaction of an incident pulse of quantum radiation and a discrete scatterer. This theory makes use of a virtual cavity mode that leaks the pulse with arbitrary number state contents, and a cascaded master equation results for the joint state of the corresponding cavity mode oscillator and the scatterer \cite{Carmichael_cas,gardiner_cas}. The interfering field from the cavity mode and the scatterer can be analyzed in terms of its eigenmodes. The eigenmodes can subsequently be addressed by including corresponding virtual pick-up cavity modes in the theory \cite{qintwpor,inoutqpulse}.  
Here, we apply this cascaded theory to the scattering of a single two-photon pulse, propagating along a  waveguide, interacting in a chiral manner with a TLS. We show that for certain pulses, the output state splits very accurately into two modes, populated by one photon each [see Fig. \ref{fig:fig1}(b)]. Such an operation, and its inverse recombination of two orthogonal modes into a single mode, may be useful for applications in quantum information science, and we discuss several properties of these processes. \\
The article is organized as follows. In Sec. \ref{sec:theory}, we review the virtual cavity approach to the preparation and detection of quantum pulses. In Sec \ref{sec:twophotonsplit}, we present a correlation function analysis of the output field and we identify the conditions for the perfect splitting of two photons in orthogonal modes. In Sec. \ref{sec:beamcombiner}, we analyze the inverse scattering process combining two orthogonal single-photon states into a single-mode two-photon state. In Sec \ref{sec:summary}, we provide a summary and outlook.

\section{Cascaded system approach to quantum pulses}\label{sec:theory}
This section reviews the virtual cavity treatment \cite{qintwpor,inoutqpulse} of quantum pulses interacting chirally with a spatially localized quantum scatterer, shown in Fig. \ref{fig:fig1}. 

\subsection{Pulses of quantum radiation} 	
	
The illustration in Fig. \ref{fig:fig1}(a), shows our physical problem of interest: a two-photon single-mode pulse scattering on a TLS in a chiral one-dimensional geometry. The photons in the single incoming temporal mode, $u(t)$, are described by the creation operator 
\begin{align}
	    \hat{b}_{u}^\dagger = \int dt\, u(t) \hat{b}^\dagger(t),
	    \label{eq:but}
\end{align}
where the bosonic creation operator, $\hat{b}^\dagger(t)$, represents the creation of a photon, passing the location of the scattering process at time $t$ \cite{Raymer_2020}.
The standard bosonic commutation relation, $[\hat{b}(t),\hat{b}^\dagger(t')]=\delta(t-t')$, in conjunction with an orthogonal set of temporal modes, $f_i(t)$, define operators on the form of Eq \eqref{eq:but} as standard bosonic operators, i.e. $[\hat{b}_i,\hat{b}_j^\dagger]=\delta_{ij}$.
The characterization of the pulse, $u(t)$, in the time domain is equivalent to the specification of a spatial wave packet, e.g, at time zero with the amplitude $u(t)$ assigned to the spatial location $x=-ct$  \cite{PRX.5.041017}. We shall without change of notation assume factorization of the mode into an oscillating phase factor with constant frequency and a slowly varying function, $u$. This is equivalent to the application of an interaction picture in which the dynamics of the resonant quantum scatterer become slow (assuming the rotating wave approximation).

\subsection{Open system treatment of the scattering of a quantum pulse}
To simulate the dynamics in Fig. \ref{fig:fig1}(a), we apply the virtual cavity approach, sketched in Fig. \ref{fig:fig1}(b), where a cavity with a time-dependent output coupling, $g_u(t)$, releases the desired pulse that subsequently scatteres on the TLS. In this approach, the light-matter interaction is represented as a direct chiral coupling of the scatterer and the virtual cavity mode.
The time-dependent coupling function, $g_u(t)$, leading to any desired temporal mode, $u(t)$, follows, e.g., by an analysis of the emission of classical fields \cite{qintwpor,gough2015generating} by
	\begin{align}
	    g_u(t) = \frac{u^*(t)}{\sqrt{1-\int^{t}_{0}dt'|u(t')|^2}}
	    \label{eq:gut}.
	\end{align}
The time evolution of the virtual cavity and the TLS is described by a Lindblad master equation in the Born-Markov approximation ($\hbar = 1$)
	\begin{align}
	    \frac{d\rho}{dt} = -i \left[\hat{H},\rho\right]+ \sum_{i=0}^n\left(\hat{L}_i\rho \hat{L}_i^\dagger - \frac{1}{2}\{\hat{L}_i^\dagger\hat{L}_i,\rho\}\right),
	    \label{eq:drhodt}
	\end{align}
where the Hamiltonian, $\hat{H}=\hat{H}_0+\hat{H}_{us}$, denotes the Hamiltonian of the $u$-cavity and the scatterer. The Hamiltonian describing the scattering writes
	\begin{align}
	    \hat{H}_{us} = & \frac{i\sqrt{\gamma}}{2} \left( g_u(t)\hata_u^\dagger \hat{\sigma}^-  - g_u^*(t)\hatsigma^+\hata_u\right),
	    \label{eq:Hus}
	\end{align}
where $\hata_u$ is the input cavity field operator and the operators $\hatsigma^+$ and $\hatsigma^-$ are the TLS raising and lowering operators. The non-interacting Hamiltonian, $\hat{H}_0$, of the two components, vanishes in the interaction picture adopted in this article. The operator, $\hat{L}_0$, represents the interference of the quantized radiation emitted  by the scatterer and the leftmost $u$-cavity along the propagation direction,
    \begin{align}
        \hat{L}_0^{(us)} = \sqrt{\gamma}\hatsigma^-+ g_u^*(t)\hata_u.
        \label{eq:L_0us}
    \end{align}
The Lindblad operators, $\hat{L}_i$, with $i=1 ... n$ represent further local dissipation mechanisms on the scatterer (vanishing in this article).
The master equation for the $u$-cavity and the TLS gives access to expectation values of the emitted field amplitude and intensity. From the quantum regression theorem \cite{Gardiner2004QuantumNA,BRE02}, we can determine the first-order correlation function, $g^{(1)}(t,t')$, and hereby a decomposition of the output field into output modes, $v_i$, with mean photon number, $n_i$, \cite{qintwpor}
	\begin{align}
	    g^{(1)}(t,t') = \expval{\hat{L}^{\mathrm{(us)}}_0(t)^\dagger\hat{L}_0^{\mathrm{(us)}}(t')} = \sum_i n_i v_i^*(t)v_i(t').
	    \label{eq:g1us}
	\end{align}

It is the quantum state contents of these output modes that are the concern of this article.

\subsection{Quantum state contents of an output wave packet}

To analyze the quantum content of any specific output mode, $v(t),$ we imagine its complete transfer into a downstream time-dependent cavity, as sketched in Fig. \ref{fig:fig1}(b). Our cascaded system is thus equipped with yet another virtual cavity. To map the full quantum content of a specific mode, $v(t)$, the cavity input coupling must be of the form  
\begin{align} \label{eq:gvt}
    g_v(t) = -\frac{v^*(t)}{\sqrt{\int_0^t dt'\,|v(t')|^2}}.
\end{align}
The master equation, Eq. \eqref{eq:drhodt}, describing the three component density matrix, $\rho_{usv}$, is equipped with the Hamiltonian
\begin{align}
    \nonumber
    H_{usv} =&  \frac{i}{2}\big(g_u(t)\sqrt{\gamma}\hata_u^\dagger\hatsigma^-+g_u(t)g_v^*(t)\hata_u^\dagger\hata_v\\
    &+\sqrt{\gamma}g_v^*(t)\hatsigma^+ \hata_v - \mathrm{H.c.}\big)
    \label{eq:H_usv}
\end{align}
and the Lindblad term
\begin{align}
    L_0^{usv} = g^*_u(t)\hata_u+\sqrt\gamma\hatsigma^-+g^*_v(t)\hata_v.
    \label{eq:L0usv}
\end{align}
To extract information about the quantum contents of an output wave packet, we solve the master equation for the full density matrix, $\rho_{usv}$, to obtain the reduced density matrix of the output field in the $v$-mode, $\rho_v = \Trace_{u,s}(\rho_{usv})$.

In the next section, we employ the outlined formalism to investigate the multimode output from two-photon scattering on a TLS. It is also possible to address the joint state of multiple output modes by introducing further cascaded virtual cavities \cite{qintwpor,inoutqpulse}. We shall, however, study the opposite situation with multiple input modes, which can be combined into a single mode by the scattering on a TLS.  


	\section{Splitting of a  single mode two photon state}\label{sec:twophotonsplit}
	
	\begin{figure}[b]
        \centering
        \includegraphics[width=1\linewidth]{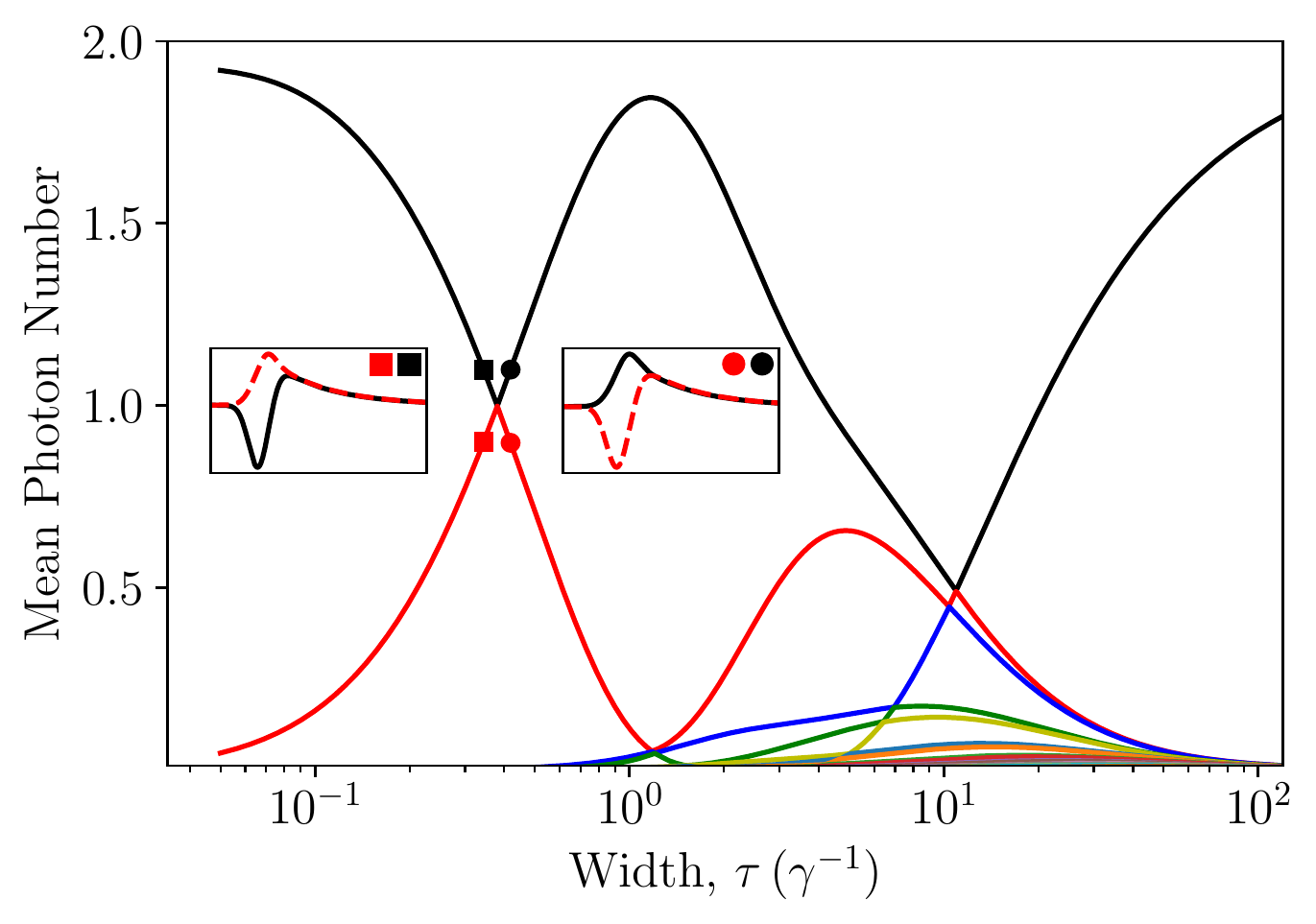}
        \caption{Scattering of a two-photon Gaussian wave packet of width, $\tau$,  on a TLS as illustrated in Fig. \ref{fig:fig1}. The average photon numbers of the ten most populated output modes are plotted as a function of $\tau$. The left and right insets show the mode shapes of the two most populated modes, $v_1$ (black full) and $v_2$ (red dashed), left (squares) and right (circles) of the crossing at $\tau\approx 0.38\gamma^{-1}$, respectively.}
        \label{fig:fig2}
    \end{figure}
    
We now consider the scattering on a TLS of two photons in a single Gaussian temporal mode, $u(t)$,
    \begin{align}
        u(t) = \frac{1}{\sqrt{\tau}\pi^{1/4}}e^{-\frac{\left(t-t_0\right)^2}{2\tau^2}}.
        \label{eq:u_gauss}
    \end{align}
    
The Hamiltonian and the emitted field operators are described in Eqs. \eqref{eq:Hus} and \eqref{eq:L_0us}, respectively. The relevant output modes are determined by the autocorrelation function  \eqref{eq:g1us}, found by the master equation \eqref{eq:drhodt} and the quantum regression theorem, (e.g., by diagonalizing the matrix representation of $g^{(1)}(t,t')$ on a two-dimensional discrete time grid). 

\subsection{Eigenmodes of the output field}

The population of the ten most populated modes is shown in Fig. \ref{fig:fig2} as a function of the input pulse duration, $\tau$. For small ($\tau \ll \gamma^{-1}$) and large durations ($\tau \gg \gamma^{-1}$) the output field predominantly occupies a  single mode. This is qualitatively expected since the short pulses are dominated by off-resonant frequency components that do not interact with the TLS, while the very long pulses have low photon coincidence probability and hence scatter in a mostly linear manner. In an intermediate pulse width regime, $\tau \lesssim \gamma^{-1}$, only two modes are significantly populated and at $\tau \approx 0.38\gamma^{-1}$ their populations cross at a value of $n_1 = n_2 = 0.9983$. The insets of Fig. \ref{fig:fig2} show the temporal mode shapes pertaining to the eigenvalues near the crossing. In particular, we observe that the modes with increasing and with decreasing population retain their shape through the crossing. The mode analysis only shows the mean occupation of the states, while the Fock state distribution can be found by solving the cascaded master equation \eqref{eq:drhodt} with the Hamiltonian  \eqref{eq:H_usv} and Lindblad operator \eqref{eq:L0usv}, as described in section \ref{sec:theory}. 

\subsection{The quantum state of the output field}

From the reduced density matrix of the downstream $v$-cavity absorbing either of the two equally populated modes, $v_1$ or $v_2$, we obtain
    \begin{align}
        \rho_{v_1} = \rho_{v_2} \approx 
        \frac{1}{2} \left(\dyad{0}{0}+\dyad{2}{2}\right)
        \label{eq:rho_v_reduced}
    \end{align}
i.e., both modes, found by diagonalization of the autocorrelation function, occupy mixed states with equal zero and two-photon components. 

The total output state is a pure state and consists almost exclusively of these two modes, which is compatible with the specific entangled state of the form
    \begin{align}
        \ket{\Psi_\mathrm{out}} = \frac{1}{\sqrt{2}}\left(\ket{0}_{v_1}\ket{2}_{v_2}-\ket{2}_{v_1}\ket{0}_{v_2}\right).
        \label{eq:bellstate}
    \end{align}
    
For two-photon scattering, the decomposition of the output wave function in two-photon states is not a coincidence: Due to bosonic statistics a two-photon wave function is symmetric concerning the change of particle indices. The wavefunction can therefore be decomposed by the so-called Takagi decomposition \cite{takagi,deterministic_photon_sorting,PhysRevA.99.023857}
    \begin{align}
        \Psi(t_1,t_2) = \sum_i \lambda_i \phi_i(t_1)\phi_i(t_2).
        \label{eq:twophotoneavefunctiontakagi}
    \end{align}
 
From the two-photon Takagi decomposed wavefunction the autocorrelation function writes
    \begin{align}
        g^{(1)}(t_1,t_2) = \sum_i |\lambda_i|^2\phi_i^*(t_1)\phi_i(t_2),
        \label{eq:g1wavefunction}
    \end{align}
which is exactly in the diagonalized form equivalent to Eq. \eqref{eq:g1us}. Hence the relation between the the mean photon numbers, $n_i$, and the singular values, $\lambda_i$, is $n_i = |\lambda_i|^2$ and the basis functions are equal up to a global phase $v_j(t) = e^{i\epsilon_j}\phi_j(t)$. Thus any output two-photon wave function will only populate the vacuum and the two-photon components when represented in the eigenbasis of the autocorrelation function, Eq. \eqref{eq:g1us}. Our calculation has identified a special case of the Takagi form with only two populated modes with equal populations. For this special case of two-fold degenerate mean photon number, the two-mode output state, Eq. \eqref{eq:bellstate}, can also be described in a rotated basis corresponding to new mode functions
    \begin{align}
        v_1'(t) &= \frac{1}{\sqrt{2}}\left[ v_1(t)+v_2(t)\right],
        \label{eq:v_1'}\\
        v_2'(t) &= \frac{1}{\sqrt{2}}\left[ v_1(t)-v_2(t)\right]
        \label{eq:v_2'}.
    \end{align}
In this basis, the output wave function decomposes into a product state of one-photon Fock states
    \begin{align}
        \ket{\Psi_\mathrm{out}} = \ket{1}_{v_1'}\ket{1}_{v_2'},
        \label{eq:psioutnewbasis}
    \end{align}
as sketched in Fig. \ref{fig:fig1}, achieving a perfect splitting of the single-mode two-photon input - the TLS operates as a ''quantum beam splitter".

\section{The inverse process: a 'beam combiner'}\label{sec:beamcombiner}

Running the process backward in time, we may apply the TLS as a \textit{beam combiner} turning two orthogonal pulses, each populated by a one-photon Fock state, into a two-photon single mode output.
To use the TLS as a beam combiner, we must scatter the time-reversed versions of the two modes in \eqref{eq:v_1'} and \eqref{eq:v_2'}
	\begin{align}
	    u_1(t) &= v_1'(-t)^*,
	    \label{eq:u_1}\\
	    u_2(t) &= v_2'(-t)^*.
	    \label{eq:u_2}
	\end{align}
The two modes will scatter into the time reverse version of the input mode \eqref{eq:u_gauss}, $v(t)=u(-t)^*$.
To demonstrate the time-reversed process, the virtual cavity approach is supplemented by another input mode. To analyze the beam combiner we shall employ an extended virtual cavity procedure  \cite{qintwpor,inoutqpulse} to deal with a multi-mode input, and we shall study the robustness of such a device to a temporal mismatch of the incident pulses.

\subsection{Two input modes and one output mode}  

 To simulate the chiral scattering of the two-mode input, we use the cascaded system consisting of two $u$-cavities, emitting the two input modes in Eq.'s \eqref{eq:u_1} and \eqref{eq:u_2}, a TLS and a single $v$-cavity, absorbing the mode, $v(t) = u(-t)^*$. The dynamics described by the Lindblad master equation with Hamiltonian 
    \begin{align}
        \nonumber
        H_{uusv} =&  \frac{i}{2}\bigg(
        g_{u_2}(t)\hata_{u_2}^\dagger\left[g^*_{u_1}(t)\hata_{u_1}
        +\sqrt{\gamma}\hatsigma^-+g^*_{v}(t)\hata_{v}\right]\\
        \nonumber
        &+g_{u_1}(t)\hata_u^\dagger\left[\sqrt{\gamma}\hatsigma^-+g^*_v(t)\hata_v\right]\\
        &+\sqrt{\gamma}g^*_v(t)\hatsigma^+\hata_v
        - \mathrm{H.c.} \bigg), 
        \label{eq:H_uusv}
    \end{align}
    and the emitted field operator 
    \begin{align}
        L_0^{uusv} = g^*_{u_2}(t)\hata_{u_2}+g^*_{u_1}(t)\hata_{u_1}+\sqrt{\gamma}\hatsigma^-+g^*_{v}(t)\hata_v.
        \label{eq:L_uusv}
    \end{align}
    
\begin{figure}[t]
        \centering
        \includegraphics[width=1\linewidth]{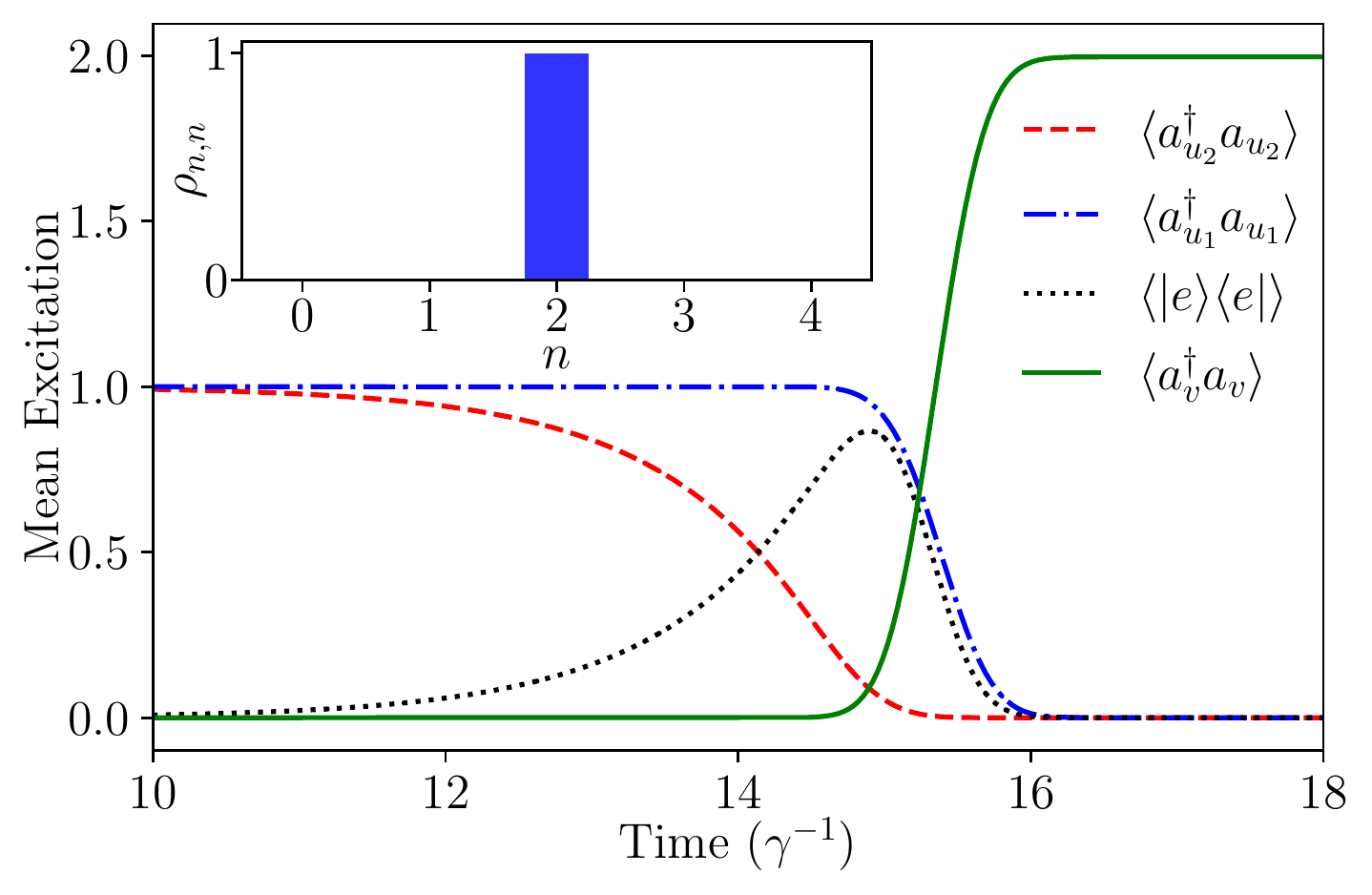}
        \caption{Two photons in different temporal modes, $u_1$ (red dashed) and $u_2$ (blue dash-dotted) given by Eq. \eqref{eq:u_1} and \eqref{eq:u_2}, are emitted from  separate virtual cavities and scattered on a TLS (black dotted). From the output, the time-reversed version of Eq. \eqref{eq:u_gauss} (green full), here called $v$, is absorbed in a virtual cavity. The mean excitation of the four system components is shown as a function of time. The two photons in their separate cavities are emitted completely, scattered on the TLS and the output cavity absorbs a single mode containing an average photon number of $\expval{\hata_v^\dagger \hata_v} = 1.999$ i.e. the full output is single mode. The inset shows the reduced density matrix elements for the output cavity for final times. Only the two-photon component is occupied with $\rho_{2,2}=0.999$. }
        \label{fig:fig3}
    \end{figure}

The coupling, $g_{u_1}(t)$, of the virtual cavity nearest to the TLS can be calculated directly using \eqref{eq:gut} with the desired mode function $u_1(t)$. The outcoupling, $g_{u_2}(t)$, of the left-most input cavity, however, must be chosen such that the dispersive scattering on the $u_1$-cavity results in the desired mode shape of $u_2(t)$. For details, see \cite{qintwpor,inoutqpulse}. 
The time-dependent population of the virtual cavities and the excited state population of the TLS are shown in Fig. \ref{fig:fig3}. The input cavity photon numbers decrease from unity to zero, while the TLS becomes temporarily excited and finally the mean photon number in the output virtual cavity mode reaches the value of two, as expected. The inset in Fig. \ref{fig:fig3}, shows the asymptotic photon number state populations in the output mode $v(t)$, in agreement with the expected pure state output. 

\subsection{Non-linear Hong-Ou-Mandel Effect}
	\label{sec:HOM}
	\begin{figure}[b]
        \centering
        \includegraphics[width=1\linewidth]{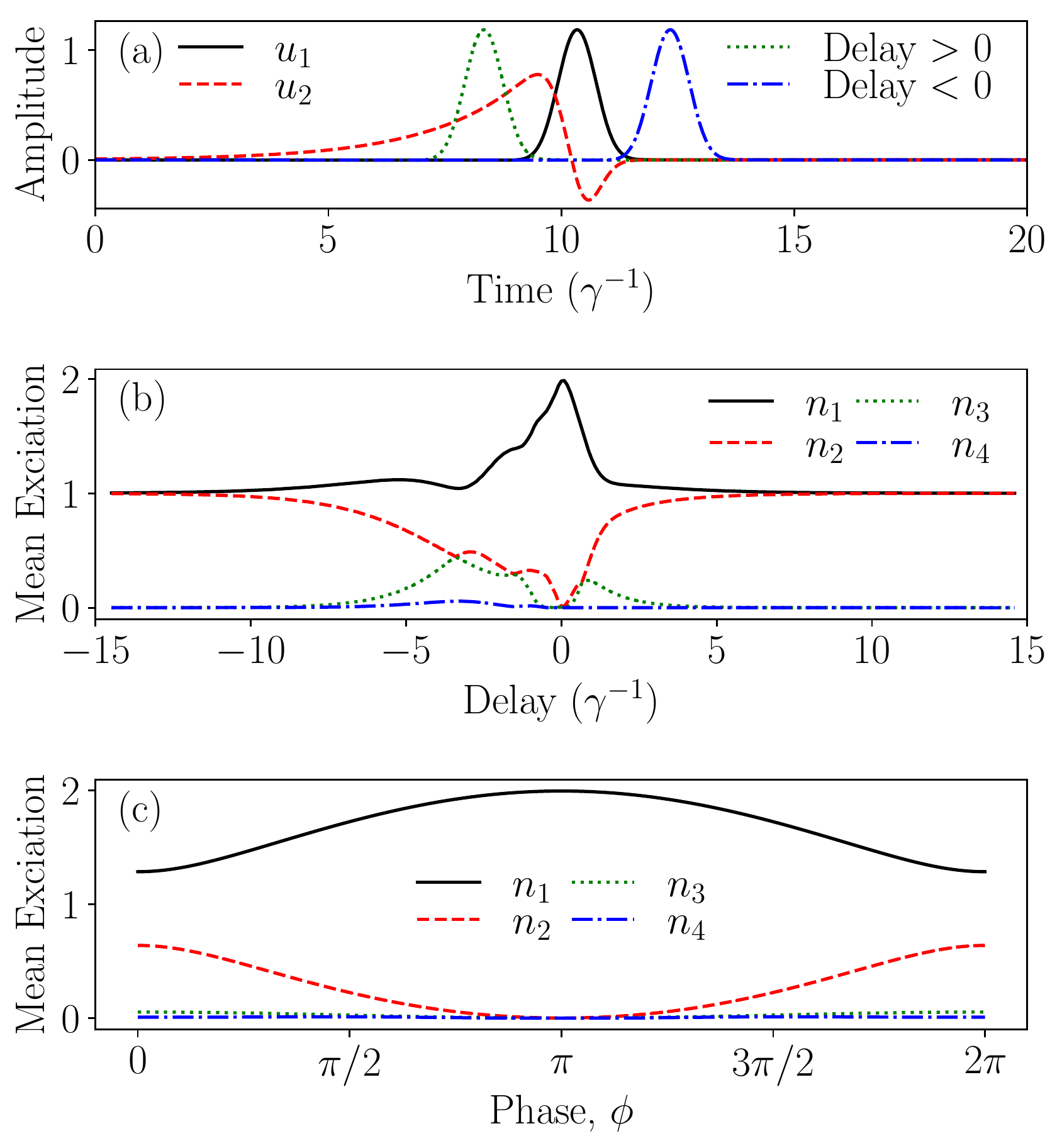}
        \caption{This figure illustrates the sensitivity of a TLS as a two-photon, two-mode beam combiner. (a) Illustration of what in section \ref{sec:HOM} is referred to as a negative (green dotted) and positive delay (blue dash-dotted) delay between the two modes $u_1$ (black full), Eq. \eqref{eq:u_1}, and $u_2$ (red dashed), Eq. \eqref{eq:u_2}. (b) Mean photon number of the four most populated output modes of the delayed two-mode two-photon input, Eq. \eqref{eq:delayedinput}, scattered on a TLS as a function of the delay between the two modes. For no delay, only one output mode is populated with two photons. For increasing the delay several modes are populated. When the modes are completely separated in time the output field consists of two modes with a single photon each. (c) The Bell state, with varying phase difference, $e^{i\phi}$, between the two states in \eqref{eq:bellstate}, is scattered on a TLS and the mean photon numbers of the output modes are shown as a function of the phase, $\phi$, for the four most populated modes. For a phase of $\phi = \pi$ only a single output mode is populated. Changing the phase from $\phi = \pi \rightarrow 2\pi$ (or $0$) a second modes becomes increasingly more populated.}
        \label{fig:fig4}
    \end{figure}

It is interesting to compare the performance of the TLS and a linear 50:50 beam splitter. When each input port of a linear 50:50 beam splitter is illuminated by a single-photon states in the same temporal mode a superposition state of both photons exiting either output port is produced. This is the famous Hong-Ou-Mandel effect \cite{HOM}, useful for testing the indistinguishably of the two incoming photons. The non-linearity of the TLS remarkably enables the beam combiner process where two photons in two orthogonal modes leave in a definite single mode, all propagating in the forward direction. The similarity and difference with the Hong-Ou-Mandel effect inspire us to ask what happens when the two photons do not arrive in the perfectly matched modes, required, e.g. one mode is delayed in time with respect to the other.
A positive delay is defined as the mode, $u_1$, arriving at the TLS at later times and a negative delay as the mode, $u_1$, arriving at the TLS at earlier times as illustrated by the dashed-dotted (blue) and dotted (green) lines in Fig. \ref{fig:fig4} (a), respectively. When delaying one of the two input modes the two modes are no longer orthogonal, but we can specify the (unnormalized) state in the second quantized form
    \begin{align}
        \ket{\Psi_\mathrm{input}} \propto \hata^\dagger_{u_1}\hata^\dagger_{u_2}\ket{0}.
        \label{eq:delayedinput}
    \end{align}
To simulate the scattering of this state on the TLS, we rewrite it in a Gram-Schmidt orthonormal basis,
    \begin{align}
    \tilde{u}_1 &= u_1
    \label{eq:u_1_tilde}\\
    \tilde{u}_2 &= \frac{1}{\sqrt{N}}\left(u_2-\langle u_1,u_2\rangle u_1\right)
    \label{eq:u_2_tilde},
    \end{align}
with the normalization factor $N=1+|\langle u_1,u_2\rangle|^2$. In the basis of the latter two modes, the input state defined by Eq. \eqref{eq:delayedinput}, up to normalization, writes
\begin{align}
        \ket{\Psi_\mathrm{input}} \propto \sqrt{N}\ket{1,1}+\langle u_1,u_2\rangle\sqrt{2}\ket{2,0}
        \label{eq:delayedinput_differentbasis}
\end{align}
where the first (second) entry is for the $\tilde{u}_{1(2)}$ mode. 
The system dynamics is now fully described by the master equation in Eq. \eqref{eq:drhodt}, with the Hamiltonian \eqref{eq:H_uusv} and Lindblad operator \eqref{eq:L_uusv}. The $u$-cavity couplings, $g_{u_i}(t)$, are defined by the $\tilde{u}_i(t)$ mode functions. 
The mean photon number in the four most populated output modes as a function of the delay between the two pulses is shown in Fig. \ref{fig:fig4}(b). For zero delay, as expected, only a single output mode is populated with two photons. For short positive delays, three modes become populated, while for longer delays two modes are populated with a single photon each corresponding to the separate linear scattering of the two incident single-photon pulses. For short negative delays, only two modes are populated while four modes are needed to account for the field at intermediate delays, and for large negative delays, we recover the separate linear scattering of two one-photon pulses.
We attribute the different features of Fig. \ref{fig:fig4}(b) between positive and negative delays to the actual mode shapes. Inspecting Fig. \ref{fig:fig4}(a), it is clear that longer negative than positive delays are needed to separate the two modes.   
Another parameter important for the TLS working as a beam combiner is the phase difference between the two states constituting Eq. \eqref{eq:bellstate}. If we replace the minus sign with a complex phase factor $e^{i\phi}$, where $\phi$ deviates from the value $\pi$, the superposition modes, $v_i'$, of Eq.'s \eqref{eq:v_1'} and \eqref{eq:v_2'}, will change and their time-reversed counterparts will not combine perfectly on the TLS. 
This is illustrated in Fig. \ref{fig:fig4}(c) which shows the average photon number, $n_i$, of the most occupied output modes as a function of the phase $\phi$. 
Changing the phase towards $\phi=0\,(2\pi$) leads to an increase in the multimode character of the output.

\section{Summary and Outlook}\label{sec:summary}
In summary, we have shown that chiral scattering of a two-photon single-mode Gaussian pulse, of a properly chosen duration, on a two-level system yields a perfect splitting into two orthogonal single-photon states. 
This result supplements the recent demonstration that a pair of two-level systems can act as a high fidelity photon sorter \cite{deterministic_photon_sorting}, which maintains the single mode character of an incident two-photon state but splits one- and two-photon components of an incident pulse in two orthogonal wave packets. 
These remarkable processes may be used in photon splitting attacks on weak pulse quantum key distribution and thus highlight the importance to develop decoy state schemes to ensure the security of quantum cryptography \cite{decoy1,decoy2,decoy3}. 
 We have demonstrated the time-reversed processes where the chiral scattering of two specific orthogonal single-photon states on a two-level systems yields a two-photon single-mode Gaussian pulse. This result may be useful to deterministically construct higher photon number states from single-photon states. 
Both results may also find applications in quantum optics and in optical quantum computing.
Recent progress with superconducting qubits coupled to microwaves \cite{microwave1,microwave2,microwave3} and surface acoustic waves \cite{phonon1,phonon2}, quantum dots chirally coupled to waveguides \cite{Lodahl2017}, and Rydberg atom clouds in free space \cite{Hofferberth1,Hofferberth2} may readily explore these processes.
Inspired by the presented results of two-photon scattering, we are inspired to extend the splitting and combing schemes to $n$-photons scattering on single or multiple emitters. We seek to realize single- or multi-photon subtraction and addition of $n$-photon input states.

\begin{acknowledgments}
This work is supported by the Danish National Research Foundation through the Center of Excellence “CCQ” (Grant agreement  No.  DNRF156) and the Carlsberg Foundation through the ``Semper Ardens'' Research Project QCooL. The authors acknowledge valuable discussions with Danil Kornovan, Anton Lauenborg Andersen, Victor Rueskov Christiansen, and Emil Vyff J{\o}rgsensen. 
\end{acknowledgments}

	\bibliography{main_txt}

\end{document}